\documentclass[twocolumn]{aastex631}

\shorttitle{Slow Rotation for Kepler-51d}
\shortauthors{Lammers \& Winn}

\graphicspath{{./}{}}
\usepackage{amsmath}
\usepackage{comment}

\begin{document}

\title{Slow Rotation for the Super-Puff Planet Kepler-51d}


\author[0000-0001-9985-0643]{Caleb Lammers}
\affiliation{Department of Astrophysical Sciences, Princeton University, 4 Ivy Lane, Princeton, NJ 08544, USA}

\author[0000-0002-4265-047X]{Joshua N.\ Winn}
\affiliation{Department of Astrophysical Sciences, Princeton University, 4 Ivy Lane, Princeton, NJ 08544, USA}

\begin{abstract}

Super-puffs are low-density planets of unknown origin and composition. If they form by accreting nebular gas through a circumplanetary disk, one might expect super-puffs to be spinning quickly. Here, we derive upper limits on the rotational oblateness of the super-puff Kepler-51d, based on precise transit observations with the NIRSpec instrument aboard the James Webb Space Telescope. The absence of detectable oblateness-related anomalies in the light curve leads to an upper limit of about $0.15$ on the planet's sky-projected oblateness. Assuming the sky-projected oblateness to be representative of the true oblateness, the rotation period of Kepler-51d is ${\gtrsim}\,40$\,hours, or equivalently, its rotation speed is ${\lesssim}\,42$\% of the breakup speed. Alternatively, if the apparently low density of Kepler-51d is due to an opaque planetary ring, the ring must be oriented within $30^\circ$ of face-on and have an inner radius smaller than $1.2$ times the planet's radius. Separately, the lack of anomalies exceeding $0.01$\% in the ingress and egress portions of the light curve places a constraint on the model of Wang \& Dai, in which the planet's apparently low density is due to a dusty outflowing atmosphere.

\end{abstract}

\section{Introduction}
\label{sec:intro}

``Super-puffs'' are exoplanets that are whimsically named but seriously perplexing \citep{Lee&Chiang2016}. The defining characteristic of a super-puff is an inferred mean density below ${\sim}\,0.1$\,g\,cm$^{-3}$, where the calculation is usually based on fitting a transit light curve to obtain the radius and analyzing transit-timing variations to obtain the mass. The known super-puffs have radii ${\gtrsim}\,5\,R_\oplus$, masses ${\lesssim}\,10\,M_\oplus$, and orbital distances ${\lesssim}\,1$\,AU. One possible structure for a super-puff is a low-mass rocky core surrounded by a hydrogen-helium envelope with a gas-to-core mass ratio of ${\sim}\,20$\,--\,$40$\% \citep{Rogers2011, Lopez2012, Batygin&Stevenson2013}. \citet{Lee&Chiang2016} proposed that such puffy planets form in the outskirts of protoplanetary disks, where it is easier for large amounts of gas to cool and accrete onto solid cores. In contrast, \citet{Millholland2019} proposed that super-puffs have gas-to-core mass ratios of $1$\,--\,$10$\%, typical of sub-Neptunes, but they are inflated by tidal heating. Others have proposed that super-puffs are not actually low-density planets: they might be planets of ordinary density that have outflowing dusty atmospheres \citep{Wang&Dai2019} or that have opaque rings viewed at low inclination \citep{Akinsanmi2020, Piro&Vissapragada2020, Ohno&Fortney2022}.

This paper is concerned with the Kepler-51 planetary system, which is blessed with (at least) three super-puffs \citep{Masuda2014}. The host is a G-type star of age ${\sim}\,500$\,Myr and distance $780$\,pc, and the three known planets have periods of about $45$, $85$, and $130$ days, placing them close to $2$:$1$ and $3$:$2$ mean-motion resonances. One might hope that having three examples in one system would help to clarify the nature of super-puffs, but so far, the system has resisted definitive interpretation. Ring systems are possible, but it seems unlikely that all three planets would be viewed at low inclination. Tidal heating seems unlikely as an explanation for Kepler-51d, in particular, because its orbital period of $130$ days and low eccentricity place the planet out of reach of the host star's tidal influence. For Kepler-51b and d, transit spectroscopy has ruled out the atmospheric scale height of several thousand kilometers that one would expect from a planet with a thick hydrogen-helium envelope \citep{LibbyRoberts2020}.

Besides transit spectroscopy, another possible probe of super-puff structure is rotation. If a super-puff grows by accreting nebular gas from the surrounding circumplanetary disk, and if its orbit is wide enough to avoid being tidally despun, one would expect it to be rotating rapidly. Furthermore, for a given rotation rate, a planet with a lower density will become more oblate, all other things being equal. In fact, the density of Kepler-51d is so low that its rotation period must be longer than about $17$\,hours to maintain hydrostatic balance.\footnote{Based on the formula $P_\mathrm{break}\,{=}\,2\pi\sqrt{R_\mathrm{eq}^3 / G M_p}$.} This can be compared with the observed rotation periods of $9.9$\,hours for Jupiter, $10.6$\,hours for Saturn, $17.2$\,hours for Uranus, and $16.1$\,hours for Neptune.

Measuring the oblateness of a transiting planet has long been appreciated as a possible avenue to learn about the planet's rotation and internal structure. The transit light curve of an oblate planet differs slightly from that of a spherical exoplanet with the same cross-sectional area \citep{Seager&Hui2002, Barnes&Fortney2003, Carter&Winn2010}. Detecting these slight differences is challenging, though. For a giant planet with a Saturn-like oblateness, one must be capable of detecting variations on the order of $100$ parts per million (ppm) or smaller that occur mainly within the relatively brief ingress and egress phases of the transit. Although this level of precision has been attained by combining multiple transit observations with the {\it Hubble}, {\it Spitzer}, and {\it Kepler} space telescopes, the planets for which the data were most precise were hot Jupiters, for which tidal despinning is expected to reduce the oblateness signal to ${\sim}\,1$\,ppm \citep{Carter&Winn2010, Zhu2014}. Here, we describe an attempt to measure the rotational oblateness of Kepler-51d, for which data from the James Webb Space Telescope ({\it JWST}) are available, for which any tidal despinning is expected to have been modest, and for which any constraints might be helpful for understanding the structure of super-puff planets.

\section{Observations}
\label{sec:observations}

The data upon which this project was based is from the {\it JWST}\ General Observer program no.\ 2571, ``Unveiling the Atmospheric Composition and Haze Formation Rates in the Young, Cool, Super-Puff Kepler-51d'' led by J.\ Libby-Roberts. Kepler-51d was observed for approximately $15$\,hours spanning a transit on June 26th, 2023, using the NIRSpec instrument. Observations were performed with the low-resolution PRISM disperser, the CLEAR filter, and the $1\farcs6$ square S1600A1 aperture, in the Bright Object Time Series mode. With these settings, the spectral resolution was about 100 and covered the wavelength range from $0.6$ to $5$\,$\mu$m. The NRS1 detector was read using the SUB512 subarray and the NRSRAPID readout pattern. A total of $18{,}082$ three-second integrations were obtained.

To extract the transit light curve from the {\it JWST} time-series data, we used the open-source \texttt{Eureka!}~pipeline \citep{Bell2022}. Our use of \texttt{Eureka!} mirrored the steps followed in other works that have processed NIRSpec data with \texttt{Eureka!} \citep[e.g.,][]{Lustig-Yaeger2023, May2023, Moran2023, Rustamkulov2023}, and we refer the reader to these works for more details about the \texttt{Eureka!}~data reduction pipeline than are given below.

Beginning with the stage 0 uncalibrated files, we performed a stage 1 reduction, using the recommended settings and an $8$-$\sigma$ cosmic ray rejection threshold. Background subtraction was performed in stage 1 by fitting a linear function to each background column in the top and bottom $7$\,pixels of the subarray. In stage 2, we ran all standard time-series observation steps, except the flat-field and photometric calibration processes, as they are unnecessary for producing a normalized light curve. Stage 3 of the \texttt{Eureka!}~pipeline carried out a second round of background subtraction and converted the 2D integration time series into 1D spectra using the optimal spectral extraction procedure of \citet{Horne1986}. During spectral extraction, we adopted an aperture half-width of $2$\,pixels and performed background subtraction with an exclusion region half-width of $3$\,pixels. In stage 4, we binned the light curve across the wavelength range $0.55$\,--\,$5.4\,\mu$m, removing outliers that were ${>}3\sigma$ from a rolling median. Because we needed to perform a custom light curve fit, we skipped stages 5 and 6 of the \texttt{Eureka!}~pipeline. The resulting light curve showed a minor downward trend with time, possibly due to starspots. Before proceeding, we flattened the light curve by fitting a linear function to the out-of-transit data and dividing the entire light curve by this best-fit linear function.

\section{Analysis}
\label{sec:analysis}

\subsection{Oblate-planet transit model}
\label{sec:oblate_model}

A planet's rotational oblateness is typically quantified by the flattening parameter,
\begin{equation}
\label{eq:f_defn}
    f = \frac{R_\mathrm{eq} - R_\mathrm{pol}}{R_\mathrm{eq}}~,
\end{equation}
where $R_\mathrm{eq}$ and $R_\mathrm{pol}$ are the equatorial and polar radii of the planet, respectively. The planet's rotational obliquity is the angle $\theta$ between the planet's polar axis and its orbital plane.

\begin{figure*}
\centering
\includegraphics[width=0.95\textwidth]{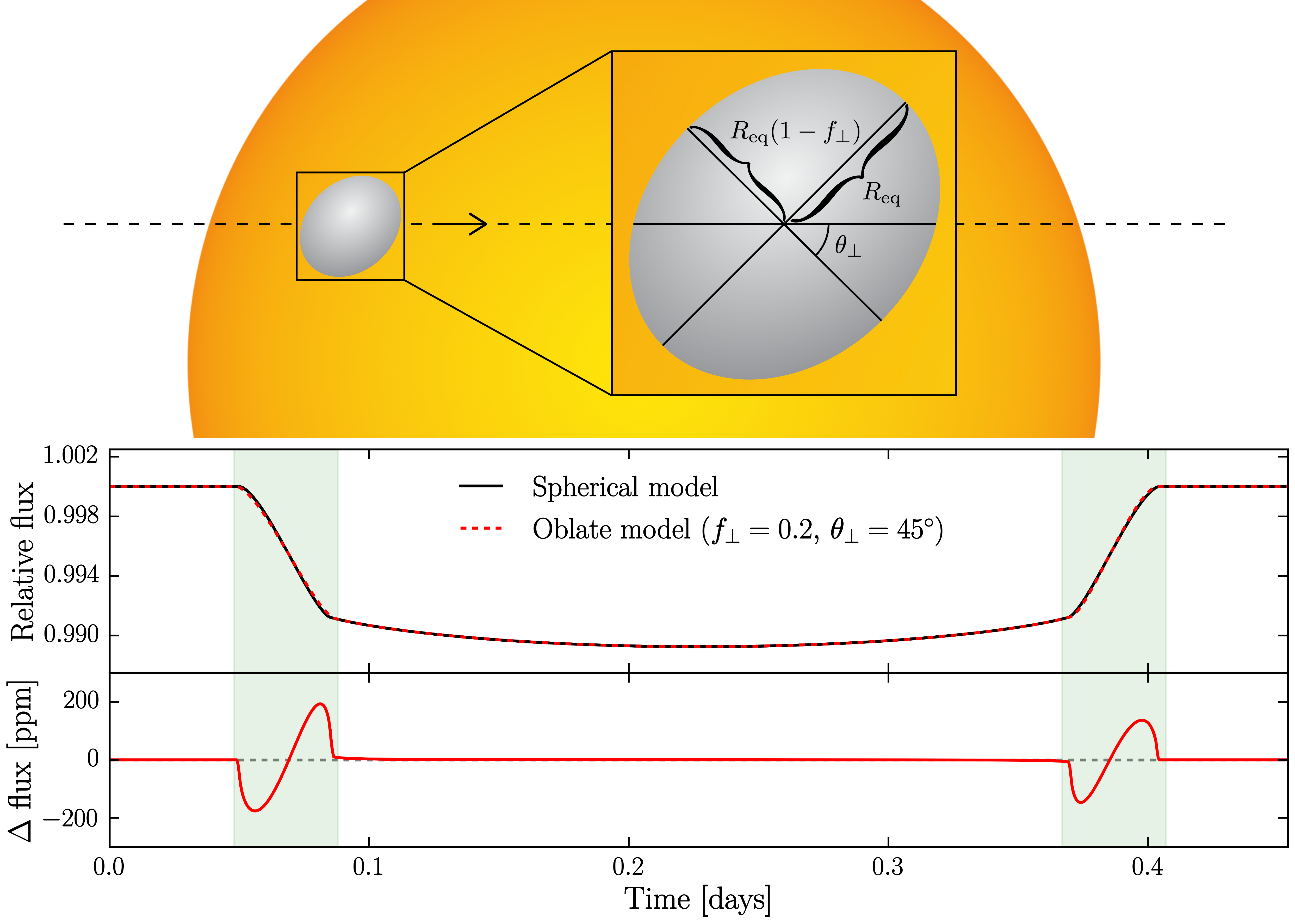}
\caption{The projected oblateness parameters (top; to-scale) and a comparison between spherical-planet and oblate-planet models for Kepler-51d (bottom). The models differ at ingress and egress (by ${\sim}\,200$\,ppm for $f_\perp\,{=}\,0.2$ and $\theta_\perp\,{=}\,45^\circ$) from that of a spherical planet with the same cross-sectional area and orbital parameters.}
\label{fig:oblate_diagram}
\end{figure*}

The transit light curve of an oblate exoplanet depends only on the planet's sky-projected shape during the transit. It is therefore convenient to parameterize the planet's shape in terms of a projected oblateness, defined as $f_\perp\,{=}\,(a\,{-}\,b)/a$, where $a$ and $b$ are the lengths of the major and minor axes, respectively, of the planet's projected ellipse. The planet's projected obliquity $\theta_\perp$ is then defined as the angle between the major axis of the projected ellipse and the transit chord. See Fig.~\ref{fig:oblate_diagram} for a diagram illustrating the geometry of an oblate transiting planet. The sky-projected oblateness and obliquity are related to $f$ and $\theta$ via the equations \citep{Barnes2009, Carter&Winn2010}
\begin{align}
\label{eq:f_theta_perp}
    f_\perp &= 1 - \sqrt{\sin^2\theta' + (1 - f)^2 \cos^2\theta'}~,\\
    \tan \theta_\perp &= \tan \theta \sin \phi~,~{\rm and} \\
    \cos^2 \theta' &= \sin^2 \theta \sin^2 \phi + \cos^2 \theta.
\end{align}
Here, $\theta'$ is the angle between the planet's spin axis and the sky plane, and $\phi$ is the azimuthal angle of the line of nodes between the planet's equatorial and orbital planes, defined such that $\phi\,{=}\,0^\circ$ when the planet's spin axis is tipped towards the observer's line of sight.

In the case where the planet's spin axis is perfectly aligned with the observer's line of sight ($\theta\,{=}\,90^\circ$ and $\phi\,{=}\,0^\circ$), the oblateness signal is undetectable because $f_\perp\,{=}\,0$ regardless of $f$. Otherwise, the strength of the oblateness signature depends on $f_\perp$, $\theta_\perp$, and the other transit parameters. Figure~\ref{fig:oblate_diagram} shows the best-fit spherical-planet model to the Kepler-51d light curve (see Section~\ref{sec:fitting}), along with an oblate-planet model with the same parameters except $f_\perp\,{=}\,0.2$ and $\theta_\perp\,{=}\,45^\circ$. In the case of Kepler-51d, these oblateness parameters result in a ${\sim}\,200$\,ppm-amplitude deviation at ingress and egress.

\begin{figure*}
\centering
\includegraphics[width=0.925\textwidth]{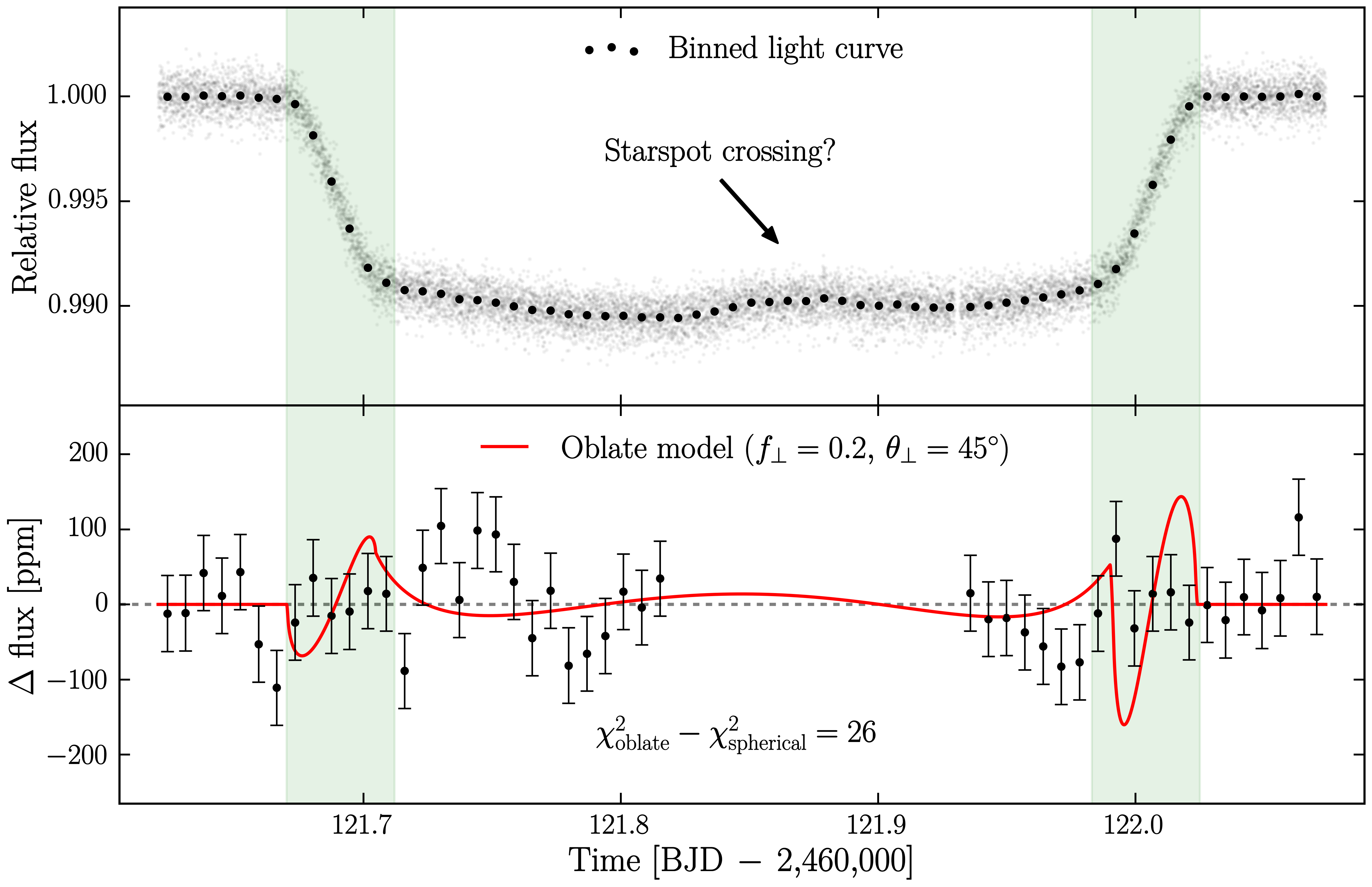}
\caption{{\it JWST} NIRSpec light curve of Kepler-51d (top) and the deviations between the data and the best-fit spherical-planet model (bottom). To illustrate the effects of oblateness on the light curve, the red curve shows the deviations between the best-fit oblate-planet model with arbitrary, fixed oblateness parameters ($f_\perp\,{=}\,0.2$ and $\theta_\perp\,{=}\,45^\circ$) and the best-fit spherical planet model. The deviations are largest during ingress and egress (highlighted in green), and are incompatible with the data ($\Delta\chi^2\,{=}\,26$). Note that the data spanning the bump in the light curve (probably due to a starspot-crossing event) were omitted before fitting.}
\label{fig:JWST_lightcurve}
\end{figure*}

\begin{figure*}
\centering
\includegraphics[width=0.775\textwidth]{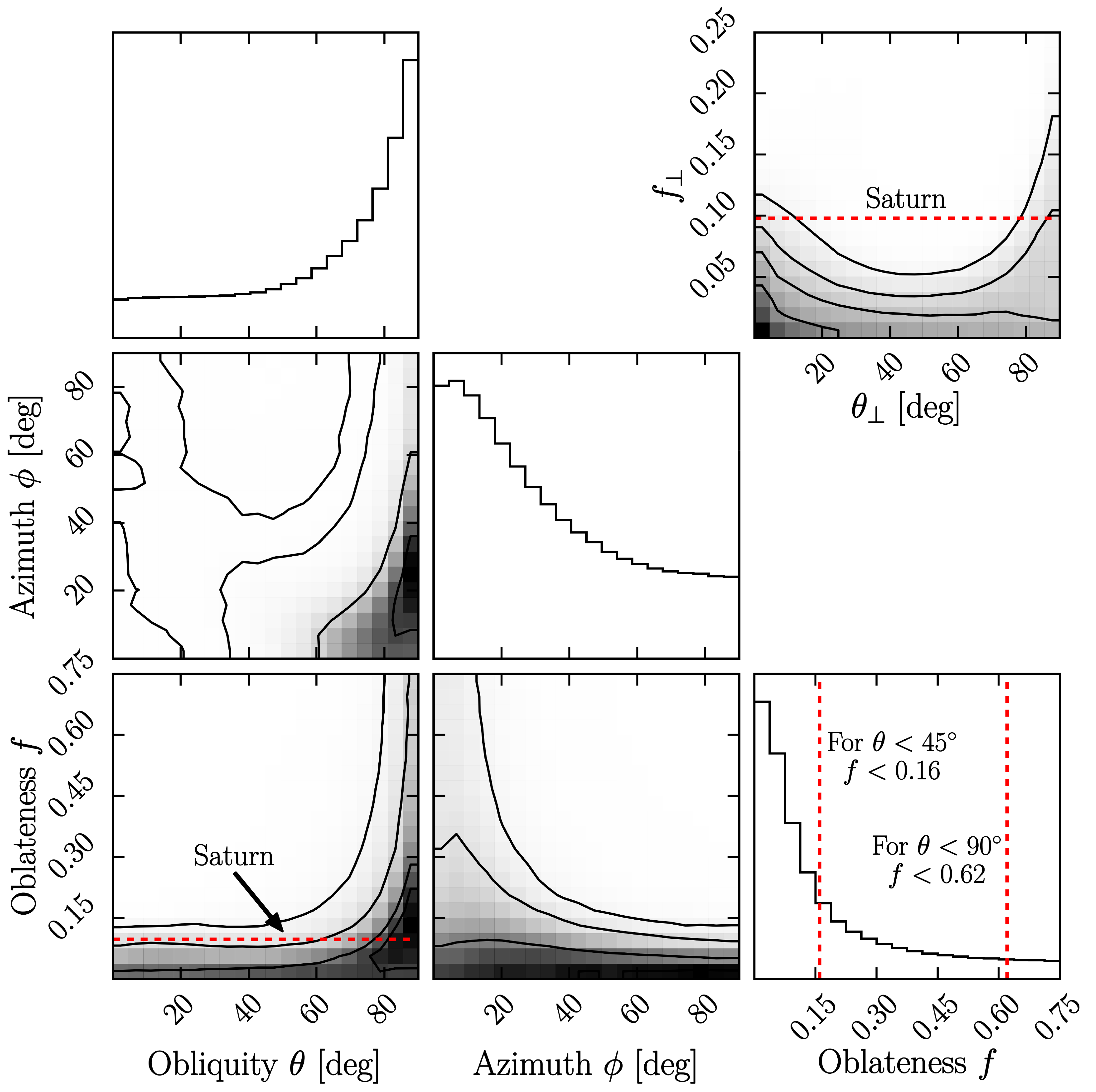}
\caption{Oblateness parameter posterior probability densities. Black contours mark the $0.5$, $1.0$, $1.5$, and $2.0$-$\sigma$ joint confidence levels, and $1D$ histograms show the marginalized distributions. Constraints on the projected quantities are shown in the top-right panel. The constraint on $f$ depends on the values of $\theta$ and $\phi$, which are unconstrained by the data. For low $\theta$, an oblateness larger than $0.15$ can be ruled out with $2$-$\sigma$ confidence, whereas for high $\theta$, $f$ is effectively unconstrained. The $95$\% confidence upper limits on $f$ for two different $\theta$ ranges are highlighted on the rightmost panel. The other parameters in the model were well-constrained without significant degeneracies; see Table~\ref{table:bestfit_vals}.}
\label{fig:MCMC_posterior}
\end{figure*}

Calculating the transit light curve is much more computationally intensive for an oblate planet than for a spherical planet. Fortunately, several authors have invested the effort to develop accurate and computationally efficient codes \citep[see, e.g.,][]{Carter&Winn2010, Liu2024}. The \texttt{JoJo} code \citep{Liu2024} and the \texttt{squishyplanet} code \citep{Cassese2024} both use Green's theorem to convert the 2D integral of the intensity of the stellar disk over the occulted area into a simpler 1D integral over the boundary. We used the \texttt{squishyplanet} code\footnote{available at \url{https://github.com/ben-cassese/squishyplanet}} to generate transit light curves. For more details on the calculation of oblate-planet transit light curves, we refer readers to the works cited above.

\subsection{Light curve fitting}
\label{sec:fitting}

Figure~\ref{fig:JWST_lightcurve} shows the light curve of Kepler-51d, consisting of $12{,}964$ measurements of the star's relative flux with a time sampling of $2.96$\,seconds. The standard deviation of the out-of-transit flux is $\sigma_\mathrm{oot}\,{=}\,715$\,ppm. For illustrative purposes, the darker circles in Fig.~\ref{fig:JWST_lightcurve} show the data after time-averaging into $65$ equally spaced bins, each consisting of approximately $209$ data points.

In addition to the usual transit features, the light curve displays a partial re-brightening event beginning near mid-transit. The duration and amplitude of the re-brightening are consistent with a starspot-crossing event, and in support of this hypothesis, we found that the amplitude of this feature decreases with wavelength. Starspot crossings were also found by \citet{Masuda2014} in {\it Kepler} data, and are expected because of the relatively young age of Kepler-51. Rather than attempting to model the starspot-crossing event, we simply masked out the region of the light curve between $\mathrm{BJD}\,{=}\,2{,}460{,}121.821$ and $2{,}460{,}121.932$, leaving $9{,}824$ data points to be fitted.

The oblate-planet transit model included the following free parameters: the mid-transit time $t_\mathrm{mid}$, the ratio of the orbital and stellar radii $a/R_\ast$, the orbital inclination $I_p$, the effective radius ratio of the planet and the star $R_\perp/R_\ast$ (where $R_\perp\,{\equiv}\,R_\mathrm{eq}\sqrt{1\,{-}\,f_\perp}$), the projected oblateness $f_\perp$, the projected obliquity $\theta_\perp$, and the re-mapped quadratic limb-darkening parameters $q_1$ and $q_2$ (as defined by \citealt{Kipping2013}). In addition, we allowed the planet's orbit to be eccentric, and thereby introduced parameters for the eccentricity $e$ and the argument of pericenter $\omega$, bringing the count to a total of ten free parameters. To set the timescale for the transit, we fixed the orbital period at the value $P\,{=}\,130.178058$\,days reported by \citet{Masuda2014}.

We applied Gaussian priors on $e\cos\omega$ and $e\sin\omega$ based on the transit-timing results reported in \citet{LibbyRoberts2020}. We also imposed a Gaussian prior on the mean stellar density, $\rho_\ast\,{=}\,2.03\,{\pm}\,0.08$\,g/cm$^3$, based on the analysis of the star's luminosity and spectroscopic properties \citep{LibbyRoberts2020}. The stellar density was related to the other parameters via the equation \citep{Winn2010, Kipping2010}
\begin{equation}
\label{eq:rho_ast}
    \rho_\ast = \frac{3\pi}{G P^2} \left(\frac{a}{R_\ast}\right)^{\!\!3} \frac{(1 - e^2)^{3/2}}{(1 + e\sin\omega)^3}~.
\end{equation}
Although the orbital eccentricity of Kepler-51d is small, we caution that fixing it to zero results in artificially strong constraints on the planet's oblateness. Uniform priors were used for all other parameters.

To prepare for a thorough investigation of the constraints on the rotational oblateness of Kepler-51d, we performed two preliminary analyses. First, we fitted the light curve assuming $f_\perp\,{=}\,0$. We minimized the usual $\chi^2$ statistic after assigning the photometric uncertainty of each flux measurement to be $\sigma_\mathrm{oot}\,{=}\,715$\,ppm. We used the Nelder-Mead optimizer implemented in the \texttt{SciPy} package's \texttt{optimize.minimize} class \citep{Virtanen2020}. The minimum $\chi^2$ was $9{,}644$ with $9{,}824$ degrees of freedom, indicating an acceptable fit. The residuals show a nearly Gaussian distribution centered on zero with no obvious patterns in time (see the bottom panel of Fig.~\ref{fig:JWST_lightcurve}). To illustrate the detectability of a hypothetical oblateness signal, we performed a second fit to the light curve in which $f_\perp$ was held fixed at $0.2$, $\theta_\perp$ was held fixed at $45^\circ$, and the other parameters were optimized. The resulting fit is not as good, with $\Delta\chi^2\,{=}\,26$ in favor of the spherical-planet model. Note that the best-fit oblateness model differs from the naive oblate-planet model shown in Fig.~\ref{fig:oblate_diagram} because the other parameters (e.g., $a/R_\ast$, $I_p$, $q_1$, $q_2$) were optimized, thereby reducing the amplitude of the deviations between the oblate-planet model and the spherical-planet model.

Second, we performed inject-and-recover experiments to ground our expectations about the achievable constraints. We created $3{,}000$ simulated light curves, each of which had the same time stamps and level of photometric noise as the real light curve, and an injected signal based on a value of $f_\perp$ drawn randomly from $0$ to $0.5$, a value of $\theta_\perp$ drawn randomly from $0$ to $90^\circ$, and other parameters taken from the best-fit spherical-planet model. In each case, we determined the posterior for $f_\perp$ using a short Monte Carlo Markov Chain (MCMC) method similar to the one described in more detail in the next section. The main results were:
\begin{itemize}
    \item When the injected signal has $f_\perp\,{\approx}\,0$, the distribution of recovered values reaches its 95th percentile level at $f_\perp\,{=}\,0.07$. Thus, we should probably not trust any claimed detection of oblateness smaller than about $0.07$.
    
    \item When the injected signal has $f_\perp\,{\gtrsim}\,0.1$, the difference between the injected and recovered value has a standard deviation of about $0.03$. This sets our expectation for the statistical uncertainty of any positive detection of oblateness.
\end{itemize}

Although we had already removed the obvious bump-like feature that we attributed to a starspot-crossing event, we remained concerned about any remaining time-correlated noise due to stellar activity or instrumental effects. To address this concern, we performed an additional experiment in which the synthetic light curves were constructed by applying random cyclic permutations of the residuals between the data and the spherical-planet model, thereby retaining the time correlations in the residuals. Repeating the MCMC analysis, we again found $f_\perp$ values above $0.07$ to be recoverable, with a standard deviation between the injected and recovered values of $0.03$.

As an aside, we also investigated the light curve of the sister planet Kepler-51b observed with {\it JWST} NIRSpec (General Observer program no.\ 2454; PI: P.\ Gao). When we reduced the data using the \texttt{Eureka!}~pipeline and repeated the preceding inject-and-recover experiments, we found the prospects for a reliable measurement of oblateness to be less promising. In this case, when the injected signal had $f_\perp\,{\approx}\,0$, the recovered values had a 95th percentile of $f_\perp\,{=}\,0.23$. Thus, we decided to focus attention on Kepler-51d.

\subsection{Constraints on oblateness}
\label{sec:MCMC}

To place quantitative constraints on the oblateness of Kepler-51d, we applied the MCMC method to the real {\it JWST} transit light curve. Although the projected quantities $f_\perp$ and $\theta_\perp$ uniquely specify the oblateness signature seen in the transit light curve, we decided to parameterize the model by the true oblateness $f$, the true obliquity $\theta$, and the azimuthal angle $\phi$. We opted for this model, despite the redundancy in the parameter space and increased computational cost, because it proved to be a convenient way to trace out the degeneracies that affect the inference of $f$ and $\theta$. All together, there were $11$ model parameters: $t_\mathrm{mid}$, $a/R_\ast$, $I_p$, $e\sin\omega$, $e\cos\omega$, $R_\perp/R_\ast$, $f$, $\theta$, $\phi$, $q_1$, and $q_2$.

We used the open-source \texttt{emcee} code of \citet{Foreman-Mackey2013}, which implements the affine-invariant MCMC sampling algorithm of \citet{Goodman&Weare2010}. We used $400$ independent walkers, each taking $60{,}000$ steps, the first $2{,}000$ of which we discarded as burn-in. The walkers were initialized based on the best-fit spherical-planet parameters, with a small amount of Gaussian noise added to ensure independence. 
The initial $\theta$ and $\phi$ values were drawn randomly from $0^\circ$ to $90^\circ$. The initial $f$ values were drawn from a uniform distribution ranging from $0$ to $0.75$. We used a wide range to fully explore the empirical constraints on the planet's shape, although we note that Maclaurin spheroids with flattening parameters as high as 0.75 are not physically realistic.\footnote{When $f$ is larger than about $0.33$, the equilibrium shape of a rotating body is not a Maclaurin spheroid but rather a Jacobi ellipsoid \citep{Press&Teukolsky1973}.} The final posterior was well-converged, with an autocorrelation length smaller than one $50$th of the total number of steps.

\begin{table}
\centering
\caption{Median values and $1$-$\sigma$ uncertainties from the MCMC fit to Kepler-51d's {\it JWST} light curve ($f$, $\theta$, and $\phi$ are omitted; see Fig.~\ref{fig:MCMC_posterior}).}
\begin{tabular}{cc}
 \hline
 Parameter & Value\\
 \hline
 $t_\mathrm{mid}$ [$\mathrm{BJD}\,{-}\,2{,}460{,}000$] & 
 $121.84739^{+0.00006}_{-0.00006}$\\
 $a_p/R_\ast$ & $^{a}124^{+1}_{-1}$\\
 $I_p$ [deg] & $89.87^{+0.02}_{-0.02}$\\
 $e\sin(\omega)$ & $^{b}0.006^{+0.007}_{-0.007}$\\
 $e\cos(\omega)$ & $^{b}0.01^{+0.01}_{-0.01}$\\
 $R_\perp/R_\ast$ & $0.0981^{+0.0002}_{-0.0002}$\\
 $q_1$ & $0.10^{+0.01}_{-0.01}$\\
 $q_2$ & $0.56^{+0.08}_{-0.07}$\\
 \hline
\end{tabular}
\tablenotetext{a}{Fit assuming a Gaussian prior on the stellar density ($\rho_\ast\,{=}\,2.03\,{\pm}\,0.08$\,g/cm$^3$) from stellar isochrones \citep{LibbyRoberts2020}.}
\tablenotetext{b}{Fit assuming Gaussian priors from analysis of the observed TTVs ($e\sin\omega\,{=}\,-0.002\,{\pm}\,0.01$ and $e\cos\omega\,{=}\,0.01\,{\pm}\,0.01$; \citealt{LibbyRoberts2020}).}
\label{table:bestfit_vals}
\end{table}

Figure~\ref{fig:MCMC_posterior} shows the posteriors for the three oblateness-related parameters, and Table~\ref{table:bestfit_vals} gives the results for the other parameters. Marginalizing over all other parameters, we found $f_\perp\,{<}\,0.14$ (at $95$\% confidence). The upper bound on $f_\perp$ is stronger if $\theta_\mathrm{\perp}\,{\approx}\,45^\circ$. The constraint on the true oblateness $f$ depends strongly on $\theta$ and $\phi$, which are largely unconstrained. Assuming the obliquity to be low, the true and projected oblateness are similar ($f_\perp\,{\approx}\,f$) and the data rule out an oblateness larger than $0.15$ with $2$-$\sigma$ confidence. However, if $\theta$ is large, it becomes possible for the planet's spin axis to be aligned with the line-of-sight ($\phi\,{\approx}\,0^\circ$), in which case the planet's projected oblateness can be much smaller than the true oblateness and $f$ is effectively unconstrained by the data. For example, applying the restriction $\theta\,{<}\,45^\circ$ leads to the constraint $f\,{<}\,0.16$ ($95$\% confidence). On the other hand, when $\theta\,{\gtrsim}\,70^\circ$, the true oblateness $f$ is hardly constrained at all.

\subsection{Constraints on rotation rate}
\label{sec:rotation_rate}

The oblateness of a rotating planet in hydrostatic equilibrium depends on the radial density distribution and compressibility of the material composing the planet. For a simple model in which the planet is an incompressible fluid, the planet takes on the shape of a \citet{Maclaurin1742} spheroid. However, gaseous planets are not well described by such a model; the effects of gravitational compression and rotationally induced compression and rarefaction cannot necessarily be neglected. The deformation of a planet's shape due to uniform rotation can be quantified by the response coefficient $\Lambda_2$, defined as \citep{dePater&Lissauer2015}
\begin{equation}
\label{eq:Lambda2_defn}
    \Lambda_2 = \frac{J_2}{q_\mathrm{r}}~,
\end{equation}
where $q_\mathrm{r}$ is the ratio of the centrifugal force to the gravitational force at the planet's equatorial radius,
\begin{equation}
\label{eq:qr_defn}
    q_\mathrm{r} = \frac{\omega^2_\mathrm{rot} R^3_\mathrm{eq}}{G M_p}~,
\end{equation}
and $J_2$ is the usual 2nd gravitational moment. For a point mass, $\Lambda_2\,{=}\,0$, whereas for an incompressible fluid, $\Lambda_2\,{=}\,0.5$. For the outer Solar System planets, $\Lambda_2$ is a measurable quantity with values ranging from $0.108$ to $0.165$ \citep{dePater&Lissauer2015}. To first-order, $f$ is related to $\Lambda_2$ according to
\begin{equation}
\label{eq:f_Lambda2}
    f = q_\mathrm{r} \left(\frac{3}{2}\Lambda_2 + \frac{1}{2}\right)~,
\end{equation}
allowing us to relate $f$ to the planetary rotation period $P_\mathrm{rot}\,{=}\,2\pi/\omega_\mathrm{rot}$. Because the true value of $R_\mathrm{eq}$ for transiting planets is unknown, and the uncertainties in $f$ and $R_\mathrm{eq}$ are strongly covariant, solving for $f$ as a function of $P_\mathrm{rot}$ requires an iterative approach. We did so by setting the well-constrained quantity $\left(R_\mathrm{eq}/R_\ast\right)^2(1\,{-}\,f_\perp)$ (essentially the transit depth) equal to the best-fit value. Then, for a given value of $P_\mathrm{rot}$, we used Newton's method to find the value of $f$ satisfying Eq.~\ref{eq:f_Lambda2}.

\begin{figure}
\centering
\includegraphics[width=0.475\textwidth]{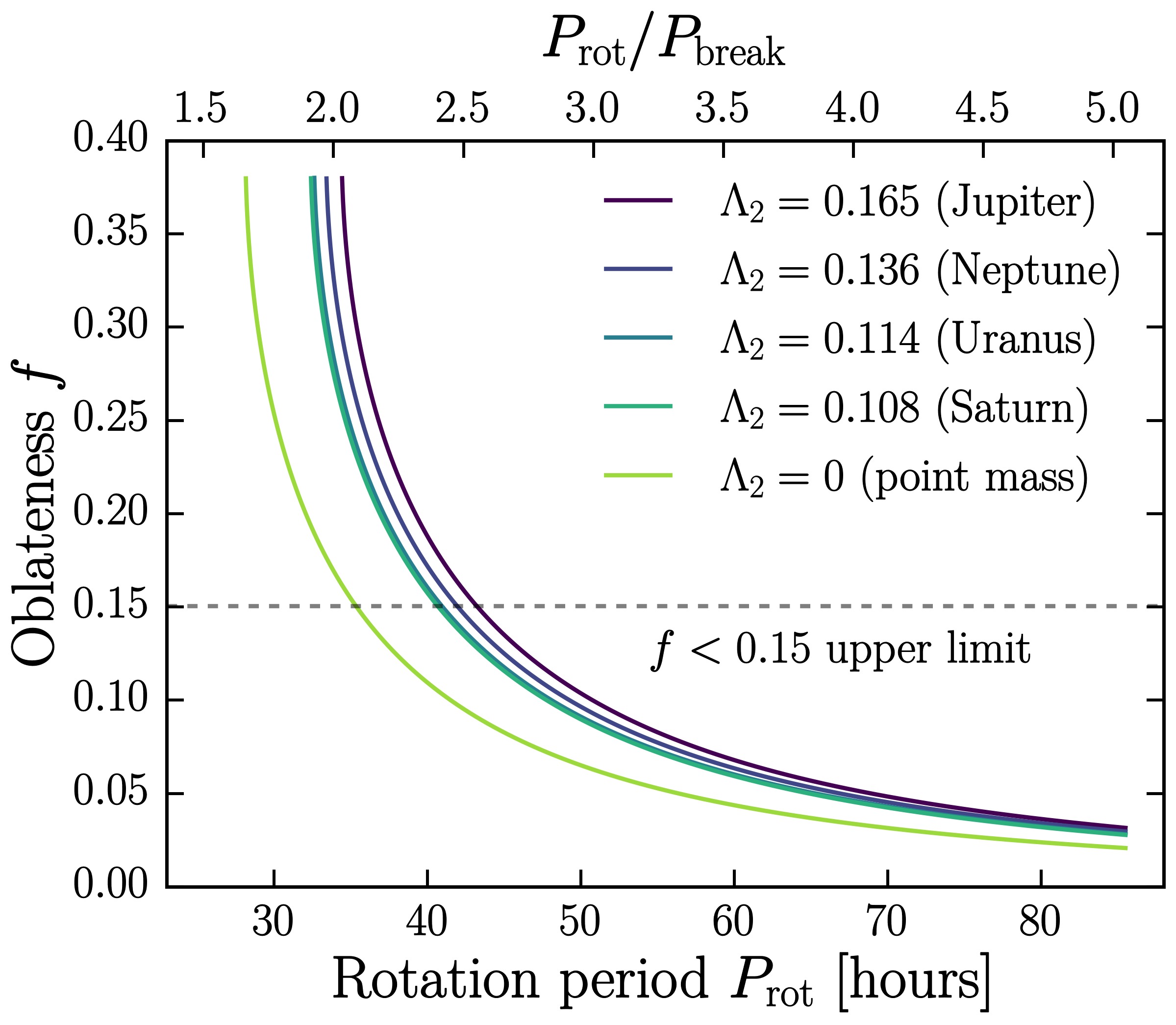}
\caption{Relationship between oblateness and rotation period for Kepler-51d with different assumed interior structures parameterized by the rotational response coefficient $\Lambda_2$. The upper $x$-axis shows the rotation period relative to the breakup period of $17$\,hours. To conform with the upper limit $f\,{<}\,0.15$, Kepler-51d requires $P_\mathrm{rot}\,{\gtrsim}\,40$\,hours, corresponding to a rotation speed ${\lesssim}\,42$\% of the breakup speed.}
\label{fig:f_vs_Prot}
\end{figure}

Figure~\ref{fig:f_vs_Prot} shows the relationship between Kepler-51d's oblateness $f$ and rotation period $P_\mathrm{rot}$ for different response coefficients. Even if Kepler-51d possess a somewhat lower $\Lambda_2$ value than the Solar System planets, it must be rotating with $P_\mathrm{rot}\,{\gtrsim}\,40$\,hours to have an oblateness consistent with $f\,{<}\,0.15$, the approximate upper limit on $f$ assuming a low obliquity. This corresponds to a rotation rate that is ${\lesssim}\,42$\% of Kepler-51d's breakup velocity.

\section{Discussion}
\label{sec:discussion}

\subsection{Comparison to other planets}

In an absolute sense, a ${\gtrsim}\,40$-hour rotation period is much longer than the $10$ to $17$-hour rotation periods of the outer Solar System planets. Relative to breakup, an upper limit on Kepler-51d's rotation rate of ${\lesssim}\,42$\% is faster than the relative rotation rates of Neptune and Uranus (both $16$\%), and approaches those of Saturn and Jupiter ($37$\% and $28$\%, respectively). Outside of the Solar System, some directly imaged planetary-mass companions (with masses ${\sim}\,10\,M_\mathrm{Jup}$) have been found to have rotation rates in the range ${\sim}\,10$\,--\,$40$\% of breakup, as inferred from rotational broadening measurements \citep[see, e.g.,][]{Snellen2014, Bryan2018, Zhou2019, Bryan2020, Zhou2020}.

The rotation rates of Uranus and Neptune are believed to have been set by the amount of gas that they accreted during formation; both have gas-to-core mass ratios of ${\sim}\,10$\% \citep{Podolak2019}. Jupiter and Saturn, on the other hand, are believed to have formed via runaway accretion, leading to primordial rotation rates close to the breakup rates. The lower rotation rates that are observed today are explained by invoking a mechanism for slowing a planet's spin shortly after formation, such as magnetic coupling with the circumplanetary disk \citep{Takata&Stevenson1996, Batygin2018, Ginzburg&Chiang2020}. The rotation rates of the directly imaged planetary-mass objects may have been set by similar mechanisms, or by processes that operate uniquely in the high-mass regime (e.g., magnetized winds). Based on its large radius and small mass, the rotation rate of Kepler-51d was likely set by the nebular gas it accreted during formation. Thus, with an inferred gas-to-core mass of ${\sim}\,40$\% \citep{Lopez&Fortney2014, LibbyRoberts2020}, it was conceivable {\it a priori} for Kepler-51d to be rotating at ${>}\,50$\% of breakup (the fastest rotating planetary-mass companion is AB Pic b, which is rotating at $67$\% of breakup; \citealt{Zhou2019}). 

\subsection{Tidal slowdown?}

Is Kepler-51d really far enough from its host star to avoid tidal despinning? The approximate spin-orbit synchronization timescale for Kepler-51d is \citep{Tremaine2023}
\begin{align}
\label{eq:tau_formula}
    \tau_s = 10.2\,\mathrm{Gyr} &\times \left(\frac{Q_p'}{10^{5.5}}\right) \left(\frac{\mathcal{C}}{0.25}\right) \left(\frac{20\mathrm{hr}}{P_\mathrm{rot}}\right) \left(\frac{M_p}{5.7M_\oplus}\right) \nonumber\\
    &\times \left(\frac{9.5R_\oplus}{R_\mathrm{eq}}\right)^{\!\!3} \left(\frac{P_\mathrm{orb}}{130.18\,\mathrm{d}}\right)^{\!\!4}~,
\end{align}
where $Q_p'$ is the planet's modified tidal quality factor and $\mathcal{C}$ is its moment of inertia divided by $M_p R_\mathrm{eq}^2$ (planetary interior models suggest that $\mathcal{C}\,\approx\,0.25$ for the outer Solar System planets; \citealt{Hubbard&Marley1989}). Tidal quality factors for the outer Solar System planets and gaseous exoplanets are estimated to be in the range $10^5$\,--\,$10^{6.5}$ \citep{Goldreich&Soter1966, Ogilvie&Lin2004, Jackson2008}. For Kepler-51d to have been tidally despun over its ${\sim}\,500$\,Myr lifetime, it must possess a low tidal quality factor of $Q_p'\,{\sim}\,10^4$.

\subsection{A high obliquity?}

Another possible explanation for the null result of our study is that Kepler-51d does have a high oblateness but is also highly oblique, with $\theta\,{\gtrsim}\,70^\circ$. It is believed that giant impacts can excite planetary obliquities to such high values \citep{Li&Lai2020, Li&Lai2021}, as proposed to explain the large obliquity of Neptune \citep{Safronov1969, Slattery1992}. However, given the near-resonant nature of the Kepler-51 planets, it seems unlikely that the system experienced any substantial dynamical scattering. In fact, motivated by the near $1$:$2$:$3$ period commensurability of the three known planets, it has been proposed that the Kepler-51 planets formed at beyond $1$\,AU in their circumplanetary disk before subsequently migrating inwards \citep{Lee&Chiang2016}. Stable high-obliquity states are known to exist for idealized resonant chain systems \citep{Millholland2024}, so if it did migrate inwards, Kepler-51d may have been captured in such a state.

\subsection{A face-on ring?}

It has also been proposed that at least some super-puffs are not oblate spheroids in hydrostatic equilibrium, as we have assumed to this point. \citet{Piro&Vissapragada2020} explored the possibility that the super-puffs are planets with extended face-on rings, which inflate their transit radii and produce flat transmission spectra. This hypothesis seems unlikely in the case of Kepler-51d because the other two known planets in the Kepler-51 system also have low densities. One would either need to invoke face-on rings for all three planets, or separate explanations for the low densities of these sibling planets. Nonetheless, the lack of an oblateness signature in Kepler-51d's {\it JWST} light curve is consistent with the face-on ring scenario. An opaque face-on ring with only a small gap between the surface of the planet and the ring's inner edge produces an indistinguishable transit light curve from a spherical planet. However, if the hypothetical ring were somewhat inclined with respect to the line of sight, or there was a significant gap between the innermost edge of the ring and the surface of the planet, an oblateness-like signal might be detectable in the transit light curve \citep{Barnes&Fortney2004}. 

To place quantitative constraints on the properties of a hypothetical ring, we carried out a similar MCMC analysis to that described in Section~\ref{sec:fitting} using the ringed-planet model implemented in the \texttt{pyPplusS} code\footnote{available at \url{https://github.com/EdanRein/pyPplusS}} \citep{Rein&Ofir2019}. Instead of using three oblateness parameters, we assumed the planet to be spherical with an opaque ring described by three parameters: the inner and outer ring radii $r_\mathrm{inner}$ and $r_\mathrm{outer}$, the inclination of the ring with respect to the line of sight $I_\mathrm{ring}$, and the sky-plane azimuthal orientation of the ring $\theta_\mathrm{ring}$.\footnote{To minimize degeneracies, the MCMC was performed with $r_\mathrm{inner}$ and the transit depth as free parameters, rather than $r_\mathrm{inner}$ and $r_\mathrm{outer}$.} We performed an MCMC analysis with $200$ walkers and $10{,}000$ steps (${>}\,15{\times}$\,the autocorrelation length), the first $1$,$000$ of which were discarded as burn-in. We were limited by the computational cost of the ring transit model. In the fit, we restricted $R_p/R_\ast$ to values between $0.02$ and $0.05$, corresponding to planet radii for which Kepler-51d's inferred density would be more typical (${\sim}\,2$\,--\,$5\,R_\earth$). We found that the following constraints can be placed on the hypothetical ring's properties: $I_\mathrm{ring}\,{<}\,30^\circ$ and $r_\mathrm{inner}\,{<}\,1.2\,R_p$ ($95$\% confidence). For comparison, Saturn's ring has an optically thick inner radius of $r_\mathrm{inner}\,{\approx}\,1.5\,R_\mathrm{Sat}$.

\subsection{A dusty outflow?}

\citet{Wang&Dai2019} put forward an alternative hypothesis to explain the flat transmission spectra of super-puffs, which involves an escaping atmosphere with high-altitude dust grains. \citet{Wang&Dai2019} predicted that the dusty outflow would manifest itself in a precise transit light curve by having less steep ingress and egress phases. Specifically, they predicted anomalies of amplitude ${\approx}\,200$\,ppm for Kepler-51b; for Kepler-51d, the signal should be larger. The {\it JWST} light curve of Kepler-51d is sufficiently precise that such a signal would be apparent in the transit light curve, but is not seen (see Fig.~\ref{fig:JWST_lightcurve}). The lack of such a feature in the light curve challenges the dusty outflow model, although the complexity of this model makes it difficult to place any quantitative constraints. Other explanations for the flat transmission spectra of Kepler-51d, such as opaque photochemical hazes \citep{Kawashima2019, Gao&Zhang2020, LibbyRoberts2020}, do not obviously predict observable anomalies in the transit light curve.

\subsection{Summary}

Measuring the oblateness of transiting exoplanets is a method by which it may be possible to constrain exoplanetary rotation rates. Kepler-51d is one of the most interesting and observationally favorable known planets for this application, given its extremely low density ($\rho_p\,{=}\,0.04$\,g\,cm$^{-3}$), deep transit (${\sim}\,1$\%), long orbital period ($P\,{=}\,130$\,days), and sufficiently bright star for precise time-series photometry with {\it JWST}. Assuming Kepler-51d is not significantly oblique ($\theta\,{\lesssim}\,45^\circ$), we found that its rotation period is likely to be longer than about $40$\,hours, or equivalently, that its rotation rate is ${\lesssim}\,42$\% of its breakup speed. As {\it JWST} turns its gaze to other far-out exoplanets, future prospects for constraining the rotation rates and obliquities of transiting exoplanets are bright.

\section{Acknowledgments} 
\label{sec:acknowledgments}

This work would not have been possible without Jessica Libby-Roberts and her co-proposers of the successful {\it JWST} program 2571, which delivered such exquisite data. We are also grateful to Ben Cassese for his \texttt{squishyplanet} code, and to Edan Rein \& Aviv Ofir for their {\tt pyPpluS} code, which accelerated our work. We would also like to thank the anonymous referee for a timely and thoughtful report, as well as Fei Dai, Akash Gupta, and Yubo Su for helpful discussions about the analysis and interpretation of our results. Lastly, we are pleased to acknowledge that the work reported in this paper was substantially performed using the Princeton Research Computing resources at Princeton University, which is a consortium of groups led by the Princeton Institute for Computational Science and Engineering (PICSciE) and the Office of Information Technology's Research Computing.

The data presented in this article were obtained from the Mikulski Archive for Space Telescopes (MAST) at the Space Telescope Science Institute. The observations can be accessed at \dataset[10.17909/yhfg-m926]{https://doi.org/10.17909/yhfg-m926}.


\bibliography{refs}{}
\bibliographystyle{aasjournal}

\end{document}